\documentclass[preprint,pra,aps,showpacs,floatfix]{revtex4}
\usepackage{amsfonts,amsmath,graphicx,latexsym}
\usepackage[utf8]{inputenc}
\usepackage{epsfig}
\usepackage{psfrag}

\newcommand{\vc}[1]{\mathbf{#1}}

\begin{document}
\title{Probing the thermal atoms of a Bose gas through Raman transition}
\author{Patrick Navez}
%
\affiliation{Katholieke Universiteit Leuven,
Celestijnlaan 200 D,
Heverlee, Belgium \\ 
Universitaet Duisburg-Essen,
Lotharstrasse 1, 47057 Duisburg,
Germany
}
\begin{abstract}
We explore the
many body physics of a Bose condensed 
atom gas at finite temperature through the Raman transition 
between two hyperfine levels. 
Unlike the Bragg scattering where the phonon-like 
nature of the collective excitations has been observed,
a different branch of thermal atom excitation is found theoretically 
in the Raman scattering.
This excitation is predicted in the generalized random phase
approximation (GRPA) and has
a gapped and parabolic
dispersion relation.
The gap energy results from the exchange interaction 
and is released during the Raman transition.
The scattering rate is determined versus the
transition frequency $\omega$ 
and the transferred momentum $\vc{q}$ and shows
the corresponding  resonance around this gap.
Nevertheless, the Raman scattering process 
is attenuated by the superfluid part of the gas. The macroscopic 
wave function of the condensate deforms its shape in order to screen 
locally   
the external potential displayed by the Raman light beams. 
This screening is total for a condensed atom transition  
in order to prevent the condensate from incoherent scattering. The 
experimental observation of this result would explain some of the reasons 
why a
superfluid condensate moves coherently
without any friction with its surrounding.
\end{abstract}
\maketitle
\section{Introduction}
\label{intro}

Among the various approximations existing in the literature 
to describe a diluted Bose 
condensed gas at finite temperature, the generalized 
random phase approximation (GRPA) has been the subject of several 
studies \cite{Reidl,Zhang,condenson,Levitov,gap}. 
This approximation has attracted a special attention 
since it is the only one in the literature with 
two important properties: 1) in agreement with the Hugenholtz-Pines 
theorem \cite{HM,HP,Ketterle,Stringari}, it predicts the observed gapless and phonon-like 
excitations; 2) the mass, momentum and energy conservation laws 
are fulfilled in the gas dynamical description. An approximation 
that satisfies these properties is said to be {\it gapless} and 
{\it conserving} \cite{Reidl,HM}.

Besides these unique features, the GRPA predicts also other 
phenomena, namely a second branch of excitations and the dynamical 
screening of the interaction potential. These phenomena appear
also in the case of a gas of charged particles or plasma. The possibility of 
a second kind of excitation has been explained quite extensively 
in \cite{condenson,Levitov,gap}. There is a distinction 
between the single particle excitations and the collective excitations.
In the case of a plasma, the first corresponds to the electrically charged 
excitations and its dispersion relation is obtained from the pole of the one 
particle Green function. The second corresponds to the plasmon
which is a chargeless excitation whose the dispersion relation is obtained 
from the pole of the susceptibility function. The plasmon mediates the interaction 
between two charged excitations. More precisely, during the interaction, one 
charged excitation emits a virtual plasmon which is subsequently reabsorbed by another 
charged excitation (see Fig.1). 
\begin{figure}
\begin{center}
\resizebox{0.50\columnwidth}{!}{\includegraphics{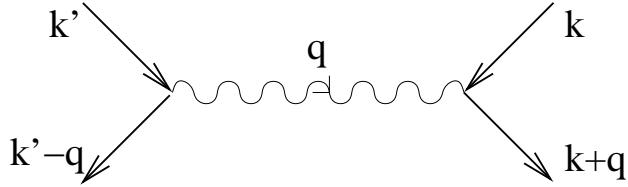}}
\end{center}

\caption{Feymann diagram illustrating the mediation process: 1) For a plasma 
two charged excitations of momentum $\vc{k}$ and $\vc{k}'$ mediate their 
interaction via a plasmon of momentum $\vc{q}$; 2) For a Bose gas, two excitations 
with one atom number unit mediate their interaction via a phonon-like collective 
excitation.}
\end{figure}

Remarkably, such a description holds also for a Bose gas with single atom excitations 
carrying 
one unit of atom number and with gapless collective excitations with no atom number.
The poles of the Green functions have a similar structure above the critical point.
But below this critical point, 
the existence of a macroscopic condensed fraction {\it hybridizes} the collective and single 
particle excitations so that 
the poles of the one particle Green function and the susceptibility function mix to form 
common branches of collective excitations \cite{gap,HM}. Thus, 
at the difference of a plasma, the presence of 
a condensed fraction prevents the direct observation of the atom-like excitation through 
the one particle Green function. 

The dynamical screening effect predicted in the GRPA appears much more spectacular in a Bose 
gas. The screening effect of the coulombian interaction 
is well known to explain the dissociation of salt diluted in water 
into its ions (see Fig.2a). 
But it also provides an explanation to the superfluidity phenomenon i.e. the 
possibility of a metastable motion without any friction. 
\begin{figure}
\begin{center}
\resizebox{0.75\columnwidth}{!}{\includegraphics{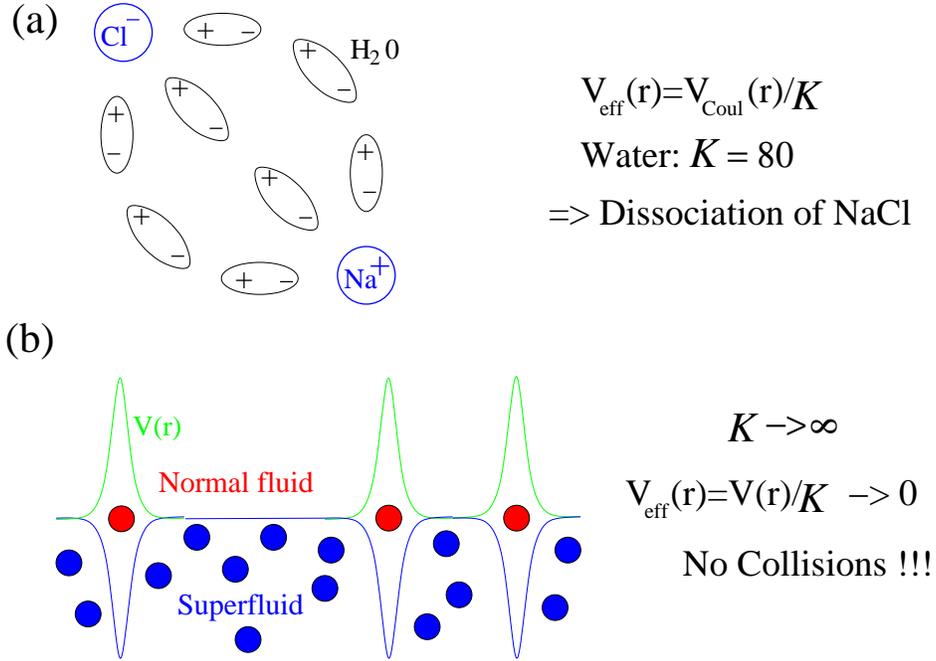}}
\end{center}
\caption{ Illustration of the screening
effect:
(a) In water, the interaction force between the ions
$Na^+$ and $Cl^-$
of the salt is screened by the presence of water molecules and
the coulombian potential $V_{Coul}(\vc{r})$ is reduced
by the relative permittivity  factor ${\cal{K}} \sim 80$.
(b)
In a Bose condensed gas, a similar effect occurs.
The condensed and thermal atoms represented in blue
and red respectively correspond  in good approximation
to the superfluid and normal fluid.
The
interaction potential $V(\vc{r})$  displayed by these thermal atoms
on condensed atoms are pictured qualitatively by the green line.
The
macroscopic wave function associated to
the condensed atoms deforms its shape in order to locally
modify the superfluid mean field interaction energy represented by the
blue line. The net result is a total screening of the interaction potential
by this mean field energy,
which prevents binary collision processes between condensed and thermal
atoms. In this way, one can explain qualitatively the metastability
of a
relative motion between the  superfluid and the normal fluid.
}
\label{fig2}

\end{figure}
Most of the literature on superfluidity
is usually devoted to the
study of metastable motion in a toroidal geometry like, for example,
an annular region between two concentric cylinders possibly
in rotation \cite{books,Leggett2}. In this simply
connected geometry, the angular momentum
about the axis of the cylinder of the superfluid
is quantized in unit of $\hbar$. The
metastability of the motion is explained by the impossibility to go
continuously from one quantized state to another due to
the difficulty to surmount an enormous free-energy barrier.
This is not the situation we want to address in this paper.
We are rather focusing on the explanation of the superfluid ability
to flow without any apparent friction with its surrounding.

The Landau criterion is a necessary but not sufficient condition for superfluidity.  It tells about the kinematic conditions under which an external object can move relatively to a superfluid without damping its relative velocity by emitting a phonon-like collective excitation. For a dilute Bose gas at low temperature, it amounts to saying that this relative velocity must be lower than the sound velocity \cite{Stringari}. The external object is assumed to be macroscopic and can be an impurity \cite{Chikkatur}, an obstacle like a lattice \cite{Cataliotti} or even the normal fluid \cite{condenson}. In particular, this criterion does not taken into account 
the fact that the normal fluid is microscopically composed of thermal excitations.  In a Bose condensed gas, even though their relative velocity is on average lower than the critical one, many of these excitations are very energetic with a relative velocity high enough to allow the phonon emission.

In the GRPA where these excitations correspond to the thermal atoms 
and under the condition of the Landau criterion, 
such a process is forbidden as shown  qualitatively from Fig.2b.
The effect of an external 
perturbation of the condensed atoms caused for example by the 
thermal atoms is attenuated by the dynamical screening. This screening 
is total in the sense that no effective mutual binary interaction 
allows a collision process which would be essential for a dissipative 
relaxation of the superfluid motion. 

The purpose of this paper is to show that these peculiar phenomena 
could in principle be observed in a Raman scattering process. 
This process induces a transition for  
a given frequency $\omega$ and a wavevector $\vc{q}$ determined from the 
difference of the frequencies and the wavevectors of 
two laser beams \cite{books}. For each wavevector corresponding to the transferred 
momentum, one can arbitrarily tune 
the frequency in order to reach the resonance energy 
associated to the 
excitation. 
Unlike the Bragg scattering which allows the observation of 
the Bogoliubov phonon-like collective excitation \cite{Ketterle,Stringari}, the Raman scattering 
is more selective. Not only the gas is probed with a selected energy 
transition and transferred momentum, but the 
atoms are scattered into a selected second internal hyperfine level.  
Through a Zeeman splitter, they can be subsequently analyzed 
separately from unscattered atoms. According to the GRPA,
the scattered thermal atoms become distinguishable from the
unscattered ones and thus release the gap energy due to the 
exchange interaction. In a previous study \cite{gap}, we showed that this 
gap appears as a resonance in the frequency spectrum of the  
atom transition rate at $\vc{q} \rightarrow 0$. The possibility 
of momentum transfer allows to analyze the influence of the screening 
of the external perturbation induced by the Raman light beams. 

The paper is divided as follows. In section 2, we review the time-dependant 
Hartree-Fock (TDHF) equations for a spinor condensate and  study  the 
linear response function to an external potential which gives results 
equivalent to the GRPA. Sections 3 and 4 are devoted 
to the Bragg and Raman scatterings respectively. Section 5
ends up with the conclusions and the perspectives.

\section{Time-dependant Hartree-Fock approximation}

We start from the time-dependant Hartree-Fock equations for 
describing two component spinor Bose gas \cite{Zhang,books} labeled by 
$a=1,2$. The atoms have a mass $m$, feel the external potential 
$V_{ab}(\vc{r},t)$ and the Hartree and Fock mean field interaction 
potential characterized by the coupling constants $g_{ab}=4\pi a_{ab}/m$ expressed in terms 
of the scattering lengths $a_{ab}$ between components $a$ and $b$ ($\hbar=1$). 
Note that no Fock mean field (or exchange) interaction 
energy appears between condensed atoms.
These 
equations describe the time evolution of a set of 
spinor wave function $\psi_{a,i}(\vc{r},t)$ describing $N_i$ atoms labeled by $i$ and depending 
on the position $\vc{r}$ and on the time $t$. 
For the condensed mode ($i=0$), these are:
\begin{eqnarray} 
\left(\begin{array}{cc}
i{\partial_t}+\frac{\nabla^2_\vc{r}}{2m}
-V_{11} & V_{12}^*\\ 
V_{12} & i{\partial_t}+\frac{\nabla^2_\vc{r}}{2m}
-V_{22}
\end{array}
\right)
\left(\begin{array}{c} \psi_{1,0} \\ \psi_{2,0}
\end{array} \right)= 
\nonumber
\\
\left(\begin{array}{cc} 
\sum_j (g_{11}
(2-\delta_{0,j})|\psi_{1,j}|^2
+g_{12}|\psi_{2,j}|^2)N_j 
&
g_{12}\sum_j (1-\delta_{0,j})N_j 
\psi_{2,j}^* \psi_{1,j}\\
g_{12}\sum_j (1-\delta_{0,j}) N_j
\psi_{1,j}^* \psi_{2,j} 
& 
\sum_j (g_{22}
(2-\delta_{0,j})|\psi_{2,j}|^2
+g_{12}|\psi_{1,j}|^2)N_j
\end{array}\right)
\left(\begin{array}{c} \psi_{1,0} \\ \psi_{2,0}
\end{array} \right)
\end{eqnarray}
For a non condensed mode ($i \not= 0$), these are
\begin{eqnarray}
\left(\begin{array}{cc}
i{\partial_t}+\frac{\nabla^2_\vc{r}}{2m}
-V_{11} & V_{12}^*\\ 
V_{12} & i{\partial_t}+\frac{\nabla^2_\vc{r}}{2m}
-V_{22}
\end{array}
\right)\left(\begin{array}{c} \psi_{1,i} \\ \psi_{2,i}
\end{array} \right)=
\nonumber
\\
\left(\begin{array}{cc} 
\sum_j (2g_{11}
|\psi_{1,j}|^2
+g_{12}|\psi_{2,j}|^2)N_j 
&
g_{12}\sum_j N_j 
\psi_{2,j}^* \psi_{1,j}\\
g_{12}\sum_j N_j
\psi_{1,j}^* \psi_{2,j} 
& 
\sum_j (2 g_{22}
|\psi_{2,j}|^2
+g_{12}|\psi_{1,j}|^2)N_j
\end{array}\right)
\left(\begin{array}{c} \psi_{1,i} \\ \psi_{2,i}
\end{array} \right)
\end{eqnarray}
The non condensed spinors remain orthogonal during their time evolution in the thermodynamic limit. 
In general, the spinor associated to the condensed mode does not remain orthogonal 
with the others. But according to  \cite{Huse}, 
the non orthogonality is not important in the thermodynamic limit for smooth external potential. 
Another way of justifying the non orthogonality is to start from an ansatz where the condensed spinor mode is described in terms of a coherent state and the non condensed ones in terms of a complete set of orthogonal Fock states i.e.
$|\Psi \rangle \sim \exp(\sum_{j\not=0} b_j c_j^\dagger-c.c.) \prod_{i\not= 0} (c_i^\dagger)^{N_i} |0\rangle$ 
where $c_i^\dagger$ is the atom creation operator in the mode $i$ and 
$b_j=\sqrt{N_0}\sum_a \int d^3 \vc{r} \psi^*_{a,j} \psi_{a,0}$.
The theory remains {\it conserving} because the conservation laws are preserved on average but 
becomes non {\it number conserving} since the quantum state is not an eigenstate of the total particle number operator. 
This procedure is justified in the thermodynamic limit since the total particle number fluctuations are relatively small during the 
time evolution.  In contrast, instead of using spinor wavefunctions, the alternative method based on the use of excitation operators is number conserving \cite{condenson,Levitov}.

The  atom number $N_i$ for each mode is supposed time-independent 
in the TDHF. Strictly speaking, a collision term must be added in 
order to allow population transfers between the various modes. 
These equations 
are valid in the collisionless regime i.e. 
on a time scale shorter than the average time 
between two collisions 
$\tau \sim 1/(\sigma_{ab} n v_T)$ where 
$v_T=\sqrt{1/\beta m}$ is the 
average velocity and $\sigma_{ab}=8\pi a_{ab}^2$ is the 
scattering cross section. In these conditions, the resulting frequency spectrum 
has a resolution limited by $\Delta \omega \sim 1/\tau$. The 
magnitude order of resolution of interest is given by the  $g_{ab}n$'s so we require $\Delta \omega /
g_{ab}n \sim \sqrt{a^3_{ab} n/\beta g_{ab}n} \ll 1$ which is generally the case when 
$a^3_{ab} n \ll 1$. These conditions are fulfilled for the parameter values considered 
in this work.

In the following, we will restrict our analysis to a bulk gas 
embedded in a volume $V$. At $t<0$, we assume all atoms in thermodynamic equilibrium in the level $1$ and 
that $V_{ab}=0$ except for $V_{22}=\omega_0$ which is 
constant and fixes the energy shift between 
the two sub-levels. In that case, the 
solutions of the TDHF are orthogonal plane waves with $i$ corresponding to the momentum  $\vc{k}$:
\begin{eqnarray}
\left(\begin{array}{c} \psi^{(0)}_{1,\vc{k}} \\ 
\psi^{(0)}_{2,\vc{k}}
\end{array} \right)=
\frac{\exp[i(\vc{k}.\vc{r}-\epsilon^{HF}_{1,\vc{k}} t)]}{\sqrt{V}} 
\left(\begin{array}{c}1
\\ 0 
\end{array} \right)
\end{eqnarray}
where we define the Hartree-Fock energy for atoms with momentum $\vc{k}$:
\begin{eqnarray}\label{drs}
\epsilon^{HF}_{1,\vc{k}}=\epsilon_\vc{k}
+g_{11} (2n- n_\vc{0}\delta_{\vc{k},\vc{0}})
\end{eqnarray}
where $\epsilon_\vc{k}=\vc{k}^2/2m$
and where the condensed and total particle densities are 
$n_\vc{0}=N_\vc{0}/V$ and $n=\sum_\vc{k} N_\vc{k}/V$. 
Eq.(\ref{drs}) corresponds to the dispersion relation 
of the single particle excitation.  
At equilibrium, 
\begin{eqnarray}
N'_\vc{k}=N_\vc{k}(1-\delta_{\vc{k},\vc{0}})=
1/(\exp[\beta(\epsilon^{HF}_{1,\vc{k}}-\mu)]-1) 
\end{eqnarray}
is the Bose-Einstein distribution. Below the condensation point,
the chemical potential becomes  
$\mu=\epsilon_\vc{0}=g_{11}(2n- n_\vc{0} )$ and the macroscopic occupation $N_\vc{0}$ is fixed 
to satisfy the total number conservation.

For $t \geq 0$, we apply 
an external potential. For the Bragg and Raman scatterings, 
these are respectively: 
\begin{eqnarray}
V_{11}=
V_B \cos(\vc{q}.\vc{r}-\omega t)
\\
V_{12}=
V_R \exp[i(\vc{q}.\vc{r}-\omega t)]
\end{eqnarray}
We solve the system through a perturbative expansion:
\begin{eqnarray}
\left(\begin{array}{c} \psi_{1,\vc{k}} \\ \psi_{2,\vc{k}}
\end{array} \right)=
\left(\begin{array}{c}e^{i(\vc{k}.\vc{r}-\epsilon^{HF}_{1,\vc{k}}t)}/\sqrt{V}
+ \psi^{(1)}_{1,\vc{k}}(\vc{r},t)
+ \psi^{(2)}_{1,\vc{k}}(\vc{r},t)
\\ \psi^{(1)}_{2,\vc{k}}(\vc{r},t)
\end{array} \right)
\end{eqnarray}
The equations of motion for the first order corrections 
are for the case of  Bragg and Raman scatterings respectively:
\begin{eqnarray}
\!\!\left[i{\partial_t}+\frac{\nabla^2_\vc{r}}{2m} -
g_{11}
(2n-\delta_{\vc{k},\vc{0}}n_\vc{0})
\right] \psi^{(1)}_{1,\vc{k}}
= \nonumber \\
\left[V_{11}+\!\sum_\vc{k'} g_{11}
(2-\delta_{\vc{k'},\vc{0}}\delta_{\vc{k},\vc{0}})
({\psi^{(0)*}_{1,\vc{k'}}} \psi_{1,\vc{k'}}^{(1)} + c.c.)
N_\vc{k'} \right]\! \psi^{(0)}_{1,\vc{k}}  \\ \label{p21}
\left[i{\partial_t}
+\frac{\nabla^2_\vc{r}}{2m}
-
\omega_0 -
g_{12}(n-\delta_{\vc{k},\vc{0}}n_{\vc{0}}) \right] \psi^{(1)}_{2,\vc{k}}
=
\left[V_{12} +
g_{12}\sum_\vc{k'}  N_\vc{k'}
{\psi^{(0)*}_{1,\vc{k'}}} \psi^{(1)}_{2,\vc{k'}}\right] 
\psi^{(0)}_{1,\vc{k}}
\end{eqnarray}

These two set of integral equations can be solved exactly using the methods 
developed in \cite{condenson}. Defining 
the Fourier transforms: 
\begin{eqnarray}
V_{ab,\vc{q},\omega}=\int_V \!\!\!d^3 \vc{r} 
\int_0^\infty \!\!\! dt\, e^{i[(\omega +i0)t -\vc{q}.\vc{r}]} V_{ab}(\vc{r},t)
\end{eqnarray}
one obtains in the level 1 for the condensed mode:
\begin{eqnarray}\label{psiB0}
\psi^{(1)}_{1,\vc{0}}(\vc{r},t)=
\sum_\vc{q'} \int_{-\infty}^\infty
\frac{d\omega'}{2\pi i}
\frac{e^{i(\vc{q'}.\vc{r}-\omega't)}V_{11,\vc{q'},\omega'}\psi^{(0)}_{1,\vc{0}}(\vc{r},t)}
{{\tilde{\cal K}}(\vc{q'},\omega')(\omega'+i0-\epsilon_{\vc{q'}})}
\end{eqnarray} 
for the non condensed modes ($\vc{k} \not=0$):
\begin{eqnarray}\label{psiB}
\psi^{(1)}_{1,\vc{k}}(\vc{r},t)=
\sum_\vc{q'} \int_{-\infty}^\infty
\frac{d\omega'}{2\pi i}
\frac{e^{i(\vc{q'}.\vc{r}-\omega't)}V_{11,\vc{q'},\omega'}\psi^{(0)}_{1,\vc{k}}(\vc{r},t)}
{{\cal K}(\vc{q'},\omega')
(\omega'+i0-\epsilon_{\vc{k}+\vc{q'}}+\epsilon_{\vc{k}})}
\end{eqnarray}
and in the level 2 for all modes:
\begin{eqnarray}\label{psiR}
\lefteqn{\psi^{(1)}_{2,\vc{k}}(\vc{r},t)=
\sum_\vc{q'} \int_{-\infty}^\infty
\frac{d\omega'}{2\pi i} \times}
\nonumber \\
& \displaystyle
\frac{e^{i(\vc{q'}.\vc{r}-\omega't)}V_{12,\vc{q'},\omega'}\psi^{(0)}_{1,\vc{k}}(\vc{r},t)}
{{\cal K}_{12}(\vc{q'},\omega')(\omega'+i0-\omega_0-\epsilon_{\vc{k}+\vc{q'}}+\epsilon_{\vc{k}}+(2g_{11}-g_{12})n
+\delta_{\vc{k},\vc{0}}(g_{12}-g_{11})n_\vc{0})}
\end{eqnarray}
These formulae resemble the one obtained from the non interacting Bose gas excepted for the mean 
field term in (\ref{psiR}) and 
the extra factors representing the screening effect. For the Bragg scattering, these factors 
can be written as \cite{condenson}: 
\begin{eqnarray}\label{Ktilde}
{\tilde{\cal K}}(\vc{q},\omega)=\frac{\Delta(\vc{q},\omega)}{(\omega+i0)^2-\epsilon_\vc{q}^2}
\\ \label{K}
{{\cal K}}(\vc{q},\omega)=\frac{\Delta(\vc{q},\omega)}{(\omega+i0)^2-\epsilon_\vc{q}^2+
2g_{11}n_\vc{0} \epsilon_\vc{q} }
\end{eqnarray} 
where 
\begin{eqnarray}
\Delta(\vc{q},\omega)=
(1-2g_{11}\chi_0(\vc{q},\omega))[(\omega+i0)^2 - {\epsilon^B_{\vc{q}}}^2]
-8g_{11}\chi_0(\vc{q},\omega)g_{11} n_{\vc{0}}\epsilon_\vc{q}
\end{eqnarray}
is the propagator for the collective excitations, 
$
\epsilon^B_{\vc{q}}=
\sqrt{c^2 \vc{q}^2 +
\epsilon_\vc{q}^2}
$ is the Bogoliubov excitation energy, $c=\sqrt{g_{11}n_\vc{0}/m}$
is the sound velocity
and 
\begin{eqnarray}\label{chi0}
\chi_{0}(\vc{q},\omega)= \frac{1}{V}\sum_{\vc{k}}
\frac{N'_{\vc{k}}-N'_{\vc{k}+\vc{q}}}
{\omega +i0 + \epsilon_{\vc{k}}-\epsilon_{\vc{k+q}}}
\end{eqnarray}
is the susceptibility function describing the normal atoms. For the Raman scattering, it is
\begin{eqnarray}\label{K12}
{{\cal K}}_{12}(\vc{q},\omega)=1-g_{12}\chi_{0,12}(\vc{q},\omega)
\end{eqnarray}
where
\begin{eqnarray}\label{chi012}
\chi_{0,12}(\vc{q},\omega)= \frac{1}{V}\sum_\vc{k}
\frac{N_{\vc{k}}}
{\omega +i0 -\omega_0 + \epsilon_{\vc{k}}-\epsilon_{\vc{k+q}}+(2g_{11}-g_{12})n+
\delta_{\vc{k},\vc{0}}(g_{12}-g_{11})n_\vc{0}}
\end{eqnarray}
Knowing the Fourier transform of the potential
$V_{11,\vc{q'},\omega'}=\sum_\pm i V_B \delta_{\vc{q'},\pm \vc{q}}/
2(\omega'+i0 \mp \omega)$ and $V_{12,\vc{q'},\omega'}= i V_R \delta_{\vc{q},\vc{q'}}/
(\omega'+i0 - \omega)$, Eqs.(\ref{psiB0},\ref{psiB},\ref{psiR}) are 
calculated using
the  contour integration method over $\omega'$ by  
analytic continuation in the lower half plane. As a consequence, the poles of the 
integrand tell about the excitation frequencies induced by the external 
perturbation. The pole of the propagator containing $\vc{k}$ corresponds to   
atom excitation involving one mode only while the poles coming from the screening 
factors correspond to the excitations involving all modes $\vc{k}$ collectively.
Thus, the TDHF approach predicts both single atom   
and collective excitations. Note that the single mode excitation is not possible 
for the condensed atoms since the corresponding pole is compensated 
by a zero coming from the screening factor. 
The expressions (\ref{psiB0},\ref{psiB},\ref{psiR}) have an 
interpretation shown in Fig.3. An 
atom of momentum $\vc{k}$ is scattered into a state of momentum $\vc{k+q'}$ 
by means of an external interaction mediated by a virtual collective excitation of momentum $\vc{q'}$.  
\begin{figure}
\label{vc}
\begin{center}
\resizebox{0.50\columnwidth}{!}
{\includegraphics{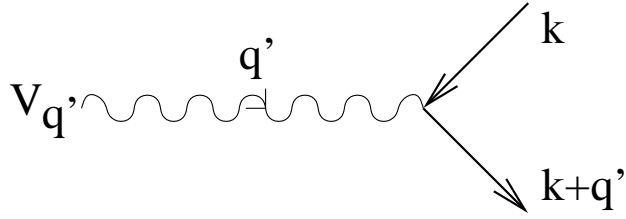}}
\end{center}
\caption{Diagrammatic representation of the scattering of an atom by an external 
potential}
\end{figure}

\section{Bragg scattering}

Let us first review the Bragg scattering process . 
Up to the second order in the Bragg potential, the atoms 
number for any mode $\vc{k}$ can be decomposed into an 
unscattered part:
\begin{eqnarray}
N_\vc{k}^{unscat}=N_\vc{k}
\left[1+\int_V d^3 \vc{r} 
(\psi^{(0)*}_{1,\vc{k}} \psi^{(2)}_{1,\vc{k}} + c.c.) 
\right]
\end{eqnarray}
and a scattered part:
\begin{eqnarray}
N_\vc{k}^{scat}=N_\vc{k}\int_V d^3 \vc{r} 
|\psi^{(1)}_{1,\vc{k}}|^2  
\end{eqnarray}
Instead of evaluating the second order term, 
$N_\vc{k}^{unscat}$ is determined through
the conservation relation $N_\vc{k}=N_\vc{k}^{unscat}+N_\vc{k}^{scat}$.
Generally speaking within the sublevel 1, the scattered 
atoms cannot be distinguished from the unscattered ones. 
But in order to understand the underlying physics, 
we assume that distinction is possible. Within the second 
order perturbation theory, the quantity of interest is 
the scattered atom rate per unit of time and is expected 
to reach a stationary value after a certain transition time. 
In the following, we shall analyze these transition rates 
for time long enough that transient effects disappear. 
In these conditions, a perturbative approach is still valid 
for very large time provided that the scattered atom number
remains low compared to unscattered ones. This last requirement is 
always satisfied with a sufficiently weak external perturbation. 

At zero temperature, only the condensed 
wave function is modified and Eq.(\ref{psiB0}) becomes 
after contour integration over $\omega'$:
\begin{eqnarray}\label{T0}
\psi^{(1)}_{1,\vc{0}}(\vc{r},t)=\frac{V_B}{2i}\psi^{(0)}_{1,\vc{0}}(\vc{r},t)
\sum_\pm e^{\pm i\vc{q}.\vc{r}}\!\!
\left[\frac{(e^{-i\epsilon^B_{\vc{q}}t}-e^{\mp i\omega t})
(\epsilon^B_{\vc{q}}+\epsilon_{\vc{q}})}{2 \epsilon^B_{\vc{q}}
(\epsilon^B_{\vc{q}} \mp \omega)}+
\frac{(e^{i\epsilon^B_{\vc{q}}t}-e^{\mp i\omega t})
(\epsilon_{\vc{q}}-\epsilon^B_{\vc{q}})}{2 \epsilon^B_{\vc{q}}
(\epsilon^B_{\vc{q}} \pm \omega)}
\right]
\end{eqnarray}
The response function is only resonant at the Bogoliubov energy $\pm \epsilon_\vc{q}^B$. 
Also no transient response appears at zero temperature. 
Using (\ref{psiB0}) and (\ref{psiB}), the total number of scattered atom 
can be obtained by determining the total momentum: 
\begin{eqnarray}\label{P}
\vc{P}=
\sum_\vc{k} N_\vc{k} \int_V d^3\vc{r}\, \psi^*_{1,\vc{k}}
\frac{\nabla_\vc{r}}{i} \psi_{1,\vc{k}}
=
\sum_\vc{k} N_\vc{k} \int_V d^3\vc{r}\, 
|\psi^{(1)}_{1,\vc{k}}|^2 \vc{q}
\end{eqnarray}
In the large time limit,
the total momentum rate 
is related to the imaginary part of the susceptibility 
response function $\chi=\chi'-i\chi''$ through \cite{Ketterle,Stringari}:
\begin{eqnarray}\label{P2}
\frac{d \vc{P}}{dt}\stackrel{t \rightarrow \infty }{=}2\vc{q} 
\left(\frac{V_B}{2}\right)^2 \chi''(\vc{q},\omega)
\end{eqnarray}
Using Eq.(\ref{T0}), we recover that: 
\begin{eqnarray}\label{suscB}
\chi''(\vc{q},\omega)=
\pi S_\vc{q}N_\vc{0}(\delta(\omega - \epsilon^B_\vc{q})-\delta(\omega +\epsilon^B_\vc{q}))
\end{eqnarray}
where $S_\vc{q}=\epsilon_\vc{q}/\epsilon^B_\vc{q}$ is the static structure 
factor. The delta function comes from the relation 
$\delta(x)=\lim_{t \rightarrow \infty} \sin(xt)/(\pi x)$.
The result (\ref{suscB}) obtained in the GRPA is identical to the one obtained from the Bogoliubov 
approach where $S_\vc{q}$ can be calculated equivalently from 
the four points correlation function \cite{Stringari,books}. But in any case the generated 
phonon like excitation is still a part of the macroscopic wave function 
$\psi_{1,\vc{0}}(\vc{r},t)$.

At temperatures different from zero, the poles become imaginary 
which means that any Bogoliubov 
excitation is absorbed by a thermal atom excitation \cite{Reidl,condenson}. 
This phenomenon is known as 
the Landau damping. So for long time, only the residues of (\ref{psiB0}) 
with poles touching the real axis contribute whereas 
the others give rise to transient terms negligible for long time. 
Thus the perturbative part becomes: 
\begin{eqnarray}
\psi^{(1)}_{1,\vc{0}}(\vc{r},t)&\stackrel{t \rightarrow \infty}{=}&
\frac{V_B}{2i}\psi^{(0)}_{1,\vc{0}}(\vc{r},t)
\sum_\pm \frac{e^{\pm i(\vc{q}.\vc{r}-\omega t)}}{
{\tilde{\cal K}}(\pm \vc{q},\pm \omega)(\pm \omega-\epsilon_\vc{q})}
\\
\psi^{(1)}_{1,\vc{k}}(\vc{r},t)&\stackrel{t \rightarrow \infty}{=}&
\frac{V_B}{2i}\psi^{(0)}_{1,\vc{k}}(\vc{r},t)
\sum_\pm \left(\frac{e^{\mp i\omega t}}{
{{\cal K}}(\pm \vc{q},\pm \omega)}-
\frac{e^{-i(\epsilon_{\vc{k}\pm \vc{q}}-\epsilon_\vc{k})t}}{
{{\cal K}}(\pm \vc{q},\epsilon_{\vc{k}\pm \vc{q}}-\epsilon_\vc{k})}\right)
\frac{e^{\pm i\vc{q}.\vc{r}}}{(\pm \omega-\epsilon_{\vc{k}\pm \vc{q}}+\epsilon_\vc{k})}
\end{eqnarray}
Using the property $\Delta(\vc{q},\omega)=\Delta^*(-\vc{q},-\omega)$, 
the total number in the condensed mode reaches a constant value
\begin{eqnarray}\label{n0scat}
N_\vc{0}^{scat}\stackrel{t \rightarrow \infty}{=}
\left(\frac{V_B}{2}\right)^2
\frac{2(\epsilon^2_\vc{q}+\omega^2)N_\vc{0}}
{|\Delta(\vc{q},\omega)|^2}
\end{eqnarray}
and the scattered thermal atom rate is given by:
\begin{eqnarray}\label{nscatt}
\frac{dN_\vc{k}^{scat}}{dt}\stackrel{t \rightarrow \infty}{=}2\pi
\left(\frac{V_B}{2}\right)^2
\sum_\pm \frac{\delta(\pm \omega-\epsilon_{\vc{k}\pm \vc{q}}+\epsilon_\vc{k})N_\vc{k}}
{|{\cal K}(\vc{q}, \omega)|^2}
\end{eqnarray}
From (\ref{P2}), we deduce for the imaginary susceptibility:
\begin{eqnarray}\label{chiT}
\chi''(\vc{q},\omega)=
-\frac{1}{g_{11}}
{\rm Im}\left(\frac{1}{{\cal K}(\vc{q},\omega)}\right)
\end{eqnarray}
The basic interpretation of these formulae is the following. At finite temperature, 
the collective excitation modes created by the external perturbation are 
damped over a time given by the inverse of the Landau damping. So the 
number of collectively excited condensed atom  
reaches the  constant value (\ref{n0scat}) 
when the produced collective excitations rate 
compensates their absorption rate by thermal atoms. This constant value is higher for 
a transition frequency and a transferred momentum close to the resonance 
$\omega=\epsilon_c \sim \pm \epsilon^B_\vc{q}$.  

The formula (\ref{nscatt}) is a generalization of the Fermi-Golden rule when 
the screening effect is taken into account. The external potential perturbs the thermal 
atoms of momentum $\vc{k}$ in two channels 
by transferring a momentum $\pm \vc{q}$ and a transition energy 
$\pm \omega$ such that the resulting single atom excitation has a momentum $\vc{k} \pm \vc{q}$ and 
a kinetic energy $\epsilon_{\vc{k} \pm \vc{q}}=\epsilon_\vc{k}\pm \omega$. 
The presence of the screening factor amplifies or reduces the scattering rate. Amplification 
(or anti-screening) occurs 
for a frequency close to the resonance energy $\epsilon_c$ of the 
collective excitations. 
On the contrary, dynamical screening occurs for a frequency close 
to the pole of the screening factor and is total 
for transition involving condensed atom at $\omega=\epsilon_\vc{q}$.
Thus, in GRPA, 
attempt to generate incoherence through single condensed atom 
scattering is forbidden at finite temperature. 
Only collective excitations affect the condensed mode but they are damped and 
therefore cannot contribute to effectively transfer condensed atoms to a different mode 
\cite{condenson}.
It is taught in standard textbooks \cite{books} that, in the impulse approximation used for large 
$\vc{q}$, the response of the system is sensitive to the momentum distribution of the gas, since the atoms behave like independent particle. In particular, a delta peak is 
expected to account for the presence of a condensate fraction. The difficulty of the observation of this peak could be explained by this impossibility of a single condensed atom excitation at finite temperature.
For completeness, let us mention that 
interaction with thermal atoms can be also totally screened and inspection of the 
formulae (\ref{K}) shows that this happens for 
$\epsilon_g=\pm \sqrt{\epsilon_\vc{q}^2-c^2\vc{q}^2}$ \cite{Zhang}. Fig. 4 shows these
features in the frequency spectrum for the total momentum rate (\ref{chiT}) at fixed $\vc{q}$. 
We choose the typical density observed experimentally for $^{87} Rb$ at the trap 
center \cite{Stringari}.
\begin{figure}
\begin{center}
\resizebox{0.75\columnwidth}{!}{
\includegraphics{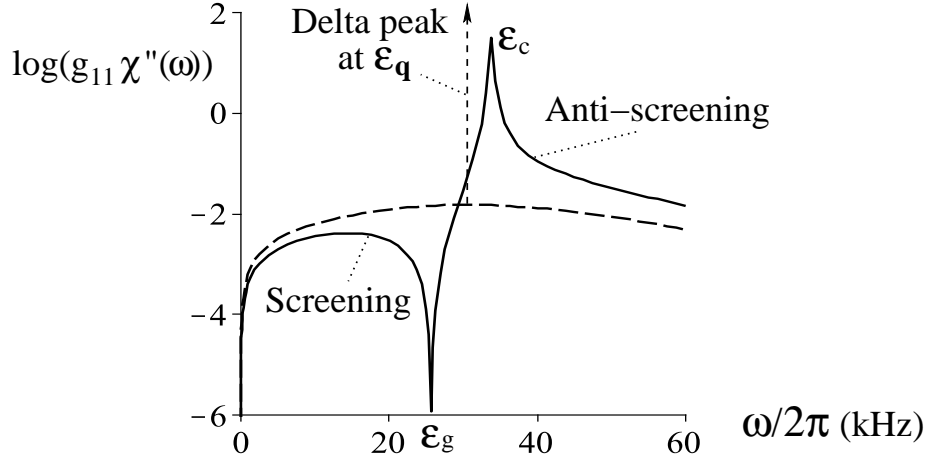}}
\end{center}
\caption{Imaginary susceptibility $\chi''$ of 
a bulk Bose condensed gas 
for 
$\epsilon_\vc{q}= 2\pi \times 30 {\rm kHz}$ versus the detuning frequency 
$\delta \omega$. 
The superfluid fraction is 94\%, 
$g_{11}n=2\pi \times 4.3 {\rm kHz}$, $k_B T/g_{11}n=2.11 $ and $a_{11}^3n= 5.6 \, 10^{-5}$.
The black dashed/solid
curve is the rate
calculated in absence/presence of the screening factor. Both regimes 
of screening and anti-screening are displayed close to the zero $\epsilon_g$ and 
to the resonance $\epsilon_c$ respectively. In particular, the screening prevents the observation 
of a huge delta peak associated to the condensed mode.}
\label{fig:0}       
\end{figure}


These results  can be put in direct relation with 
the analysis of impurity scattering \cite{Nozieres}. Indeed,
the dynamic response function is related to the dynamic 
structure factor through the fluctuation-dissipation theorem: 
$S(\vc{q},\omega)=\chi''(\vc{q},\omega)/\pi(1-\exp(-\beta \omega))$. 
The dynamic  structure factor is directly connected to the 
transition probability rate ${\cal P}(\vc{q},\omega)$ 
that an external particle or impurity 
changes its initial momentum $\vc{p}$ and energy $E_\vc{p}$ into  
$\vc{p} +\vc{q}$ and $E_{\vc{p}+\vc{q}}=E_\vc{p}+\omega$ respectively:
\begin{eqnarray}
{\cal P}(\vc{q},\omega)=2\pi |{\cal{V}}_\vc{q}|^2 S(\vc{q},\omega)
\end{eqnarray}
where $\cal{V}_\vc{q}$ is the Fourier transform of the 
interaction potential between the impurity and the atom gas.
The total rate of scattering $\Gamma_\vc{p}$ results from a virtual process involving emission and absorption of the collective excitations:
\begin{eqnarray}\label{Gam}
\Gamma_\vc{p}&=&\sum_\vc{q} 
2\pi |{\cal{V}}_\vc{q}|^2 S(\vc{q},E_{\vc{p}+\vc{q}}-E_\vc{p})
\\ \label{Gam2}
&=&\sum_{\vc{q},\vc{k}} 
2\pi |\frac{{\cal{V}}_\vc{q}}{{\cal{K}}(\vc{q},E_{\vc{p}+\vc{q}}-E_\vc{p})}|^2
\delta(\epsilon_\vc{k}+E_\vc{p}-E_{\vc{p}+\vc{q}}-
\epsilon_{\vc{k}-\vc{q}})N'_\vc{k}(1+N'_{\vc{k}-\vc{q}})
\end{eqnarray}
As a consequence, the impurity scattering is possible provided 
that the energy and momentum are  conserved in a effective collision 
with a thermal atom of momentum 
$\vc{k}$ mediated by 
a virtual collective excitation. Note that total screening prevents 
impurity scattering involving ongoing and outgoing condensed atoms. 
In contrast, for temperature close to zero, the Landau damping 
approaches zero since $\chi_0(\vc{q},\omega) \rightarrow 0$ 
so that the application of Eq.(\ref{suscB}) to (\ref{Gam})
leads to an on-energy shell process of absorption and emission of a collective 
excitation. 
We obtain:
\begin{eqnarray}\label{Gam3}
\Gamma_\vc{p}&=&\sum_{\pm,\vc{q}} 
2\pi |{\cal{V}}_\vc{q}|^2 S_\vc{q} 
(n^B_{\vc{q}}+\delta_{\pm,+})
\delta(\pm \epsilon^B_\vc{q} +E_{\vc{p}+\vc{q}}-E_\vc{p})
\end{eqnarray}
where $n^B_{\vc{q}}=1/(\exp(\beta \epsilon^B_\vc{q})-1)$.
This limit case leads to the apparent interpretation of an 
impurity interacting with a 
thermal bath of phonon-like quasi-particle.
This situation has been considered in \cite{Montina} in the study of 
the impurity dynamics. 
Instead, Eq.(\ref{Gam2}) provides a  generalization for 
higher temperature emphasizing that any external particle can excite 
a single thermal atom alone but not a condensed one.

\section{Raman scattering}

The conclusions so far obtained in the Bragg process can be extended 
straightforwardly to the 
case of Raman scattering with the difference that only one channel of scattering 
is possible. For the purpose of simplicity, we choose the case $g=g_{ab}$. 
Also this channel is easier to access experimentally. 
Defining the detuning $\delta \omega=\omega-\omega_0$, 
explicit calculations of the spinor component  (\ref{psiR})
in the second sublevel give:
\begin{eqnarray}
\psi^{(1)}_{2,\vc{k}}(\vc{r},t)&\stackrel{t \rightarrow \infty}{=}&
\left(\frac{e^{- i\omega t}}{
{{\cal K}}_{12}(\vc{q},\omega)}-
\frac{e^{i(\epsilon_\vc{k}+gn-\epsilon_{\vc{k}+\vc{q}}-\omega_0)t}}{
{{\cal K}}_{12}(\vc{q},\omega_0+ \epsilon_{\vc{k} + \vc{q}}-\epsilon_\vc{k}-gn)}\right)
\frac{e^{i\vc{q}.\vc{r}}V_R\psi^{(0)}_{1,\vc{k}}(\vc{r},t)}
{i(\delta\omega+\epsilon_\vc{k}+gn-\epsilon_{\vc{k}+\vc{q}})}
\end{eqnarray}
So we obtain for the atom number in the mode $\vc{k}$:
\begin{eqnarray}
\frac{dN_{2,\vc{k}}}{dt}\stackrel{t \rightarrow \infty}{=}2\pi
V_R^2
\frac{\delta(\delta \omega-\epsilon_{\vc{k}+\vc{q}}+\epsilon_\vc{k}+gn)N_\vc{k}}
{|{\cal K}_{12}(\vc{q},\omega)|^2}
\end{eqnarray}
By summing over all the modes, we obtain the density rate transferred in level 2:
\begin{eqnarray}
\frac{dn_{2}}{dt}=\frac{d}{dt}(\sum_\vc{k} N_{2,\vc{k}}/V)
\stackrel{t \rightarrow \infty}{=}2
V_R^2 \chi''_{12}(\vc{q},\omega)
\end{eqnarray}
where we define the imaginary part $\chi_{12}=\chi'_{12}-i\chi''_{12}$ of the 
intercomponent susceptibility function:
\begin{eqnarray}\label{chiRPA}
\chi_{12}(\vc{q},\omega)=
\chi_{0,12}(\vc{q},\omega)/(1-g\chi_{0,12}(\vc{q},\omega))
\end{eqnarray}
This last formulae is also the one obtained in the GRPA \cite{Levitov}. Again we 
find a similar structure as the intracomponent case. 
In this process, thermal atoms with an initial momentum $\vc{k}$
and energy $\epsilon^{HF}_{1,\vc{k}}= 2gn + \epsilon_\vc{k}$
are transferred into a second level with momentum
$\vc{k+q}$ and energy $\epsilon^{HF}_{2,\vc{k+q}}=gn
+\epsilon_{\vc{k}+\vc{q}}$ provided 
$\delta \omega= \epsilon^{HF}_{2,\vc{k+q}}- \epsilon^{HF}_{1,\vc{k}}$. 
In absence of screening,
a resonance appears at the detuning $\epsilon_g=\epsilon_\vc{q}-gn$.
The first term corresponds to the usual recoil energy
while the second is the gap energy $gn$ that 
results from the exchange interaction.
During the Raman transition, the transferred atoms 
become distinguishable from the others and 
release this gap energy. 

\begin{figure}
\resizebox{1\columnwidth}{!}{
\includegraphics{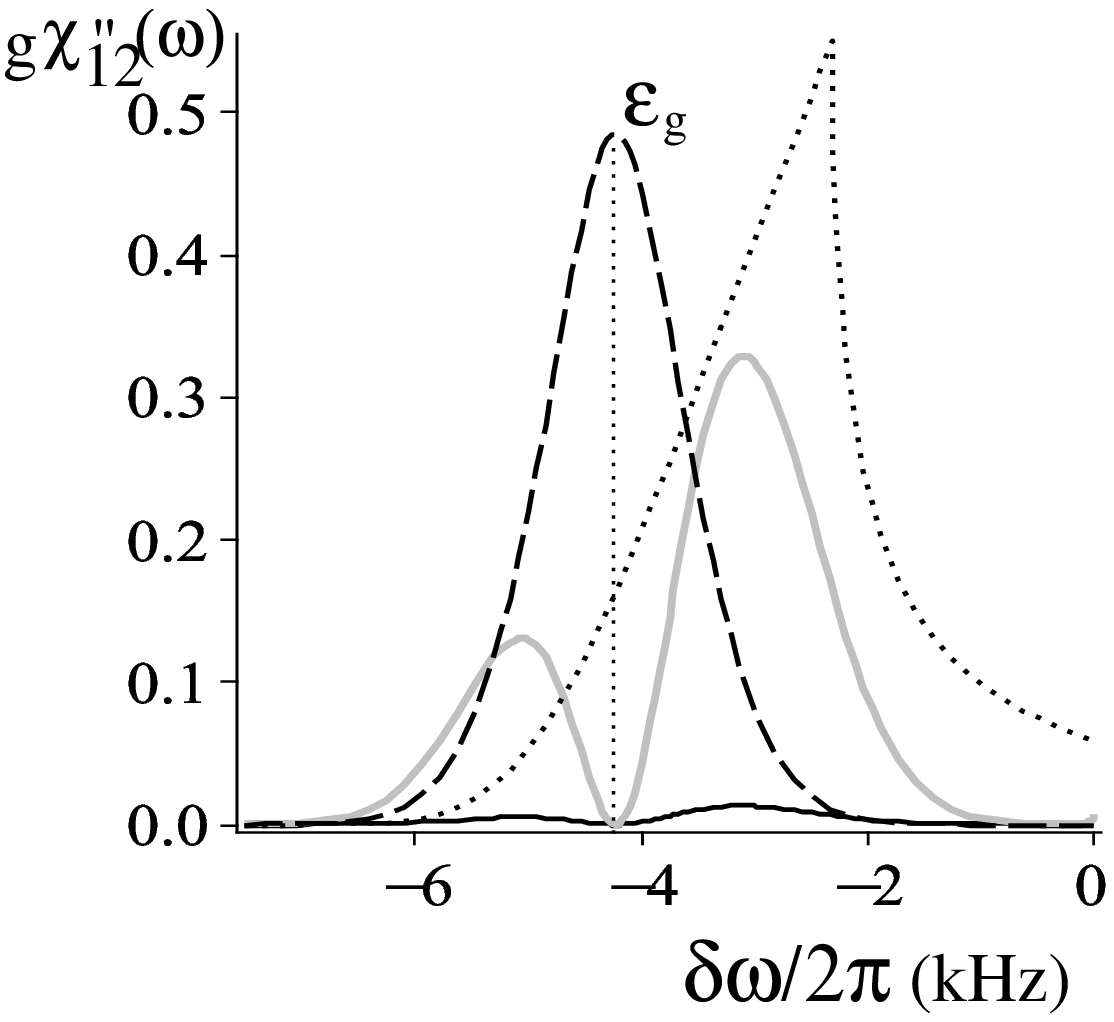} \includegraphics{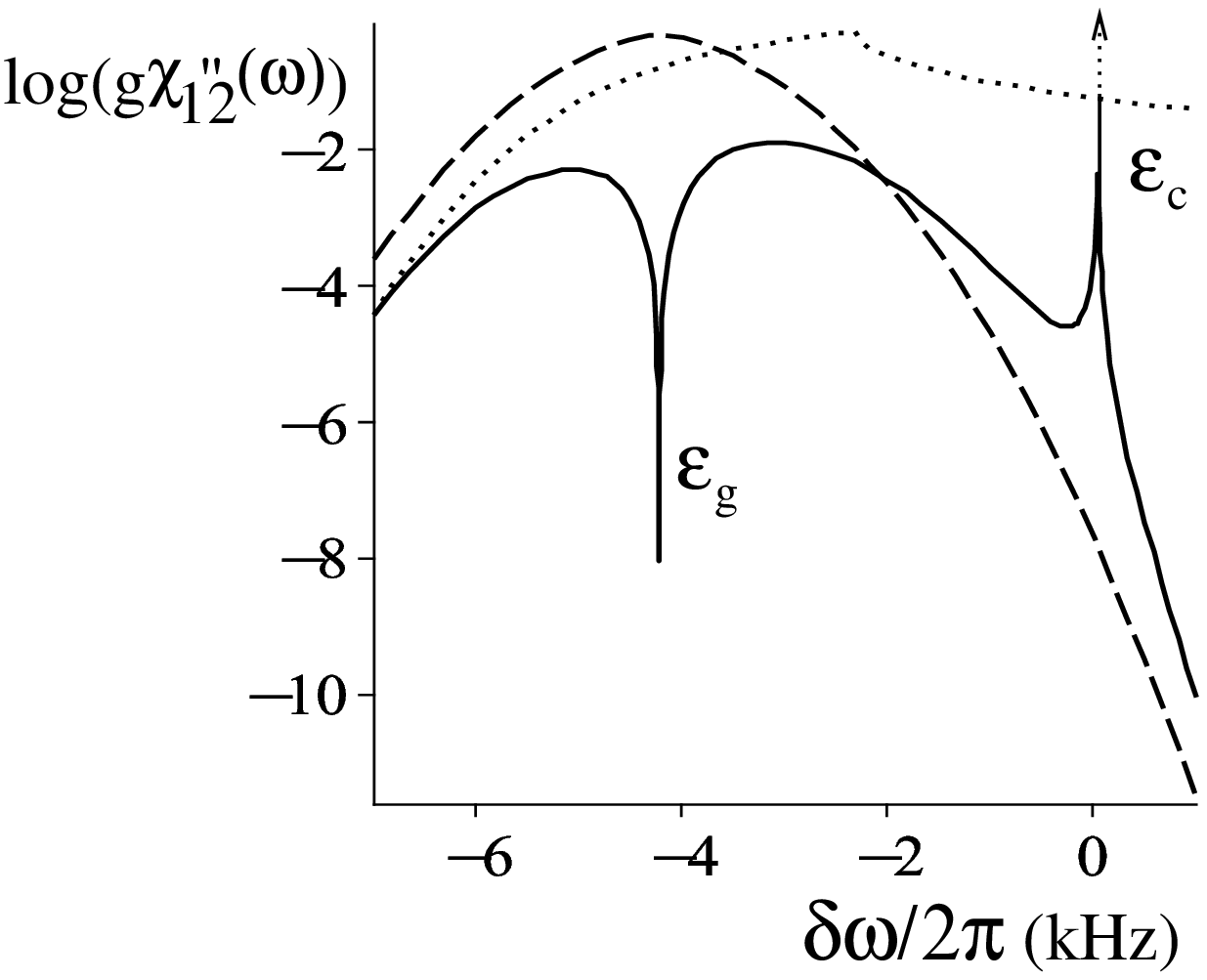}}
\caption{Imaginary susceptibility $\chi''_{12} $ of 
a bulk Bose condensed gas 
$\epsilon_\vc{q}= 2\pi \times 30 {\rm Hz}$ versus the detuning frequency 
$\delta \omega$. Parameter values are the ones of Fig.4.
Left and right graphs represent the same 
curves but the right graph is in logarithm scale. The black dashed/solid
curve is 
calculated in absence/presence of the screening factor while the dotted 
curve represents the Bogoliubov approximation. See the grey curve for
a magnification of the
black solid curve ($\times 25$) }
\label{fig:1}       
\end{figure}
\begin{figure}
\resizebox{1\columnwidth}{!}{
\includegraphics{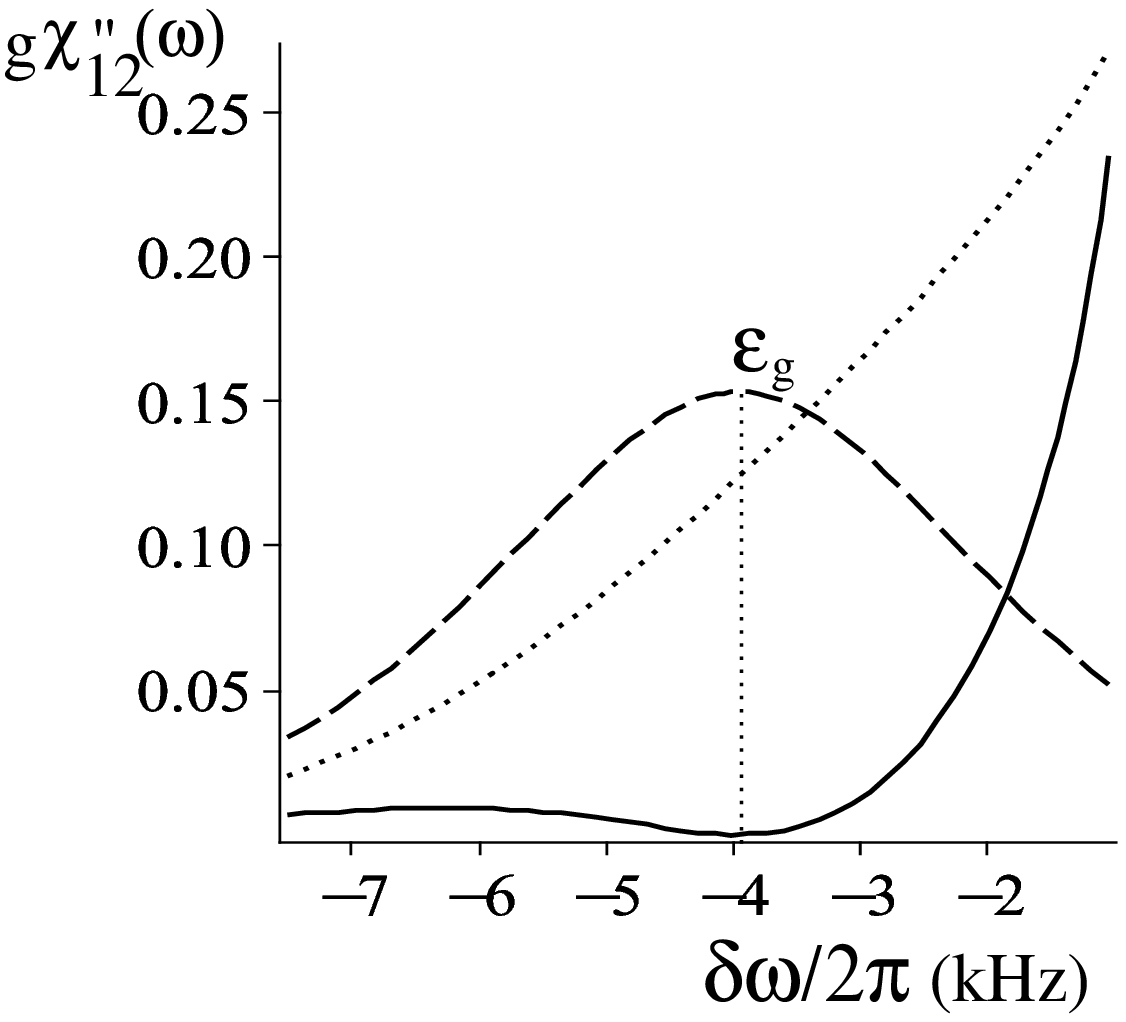} \includegraphics{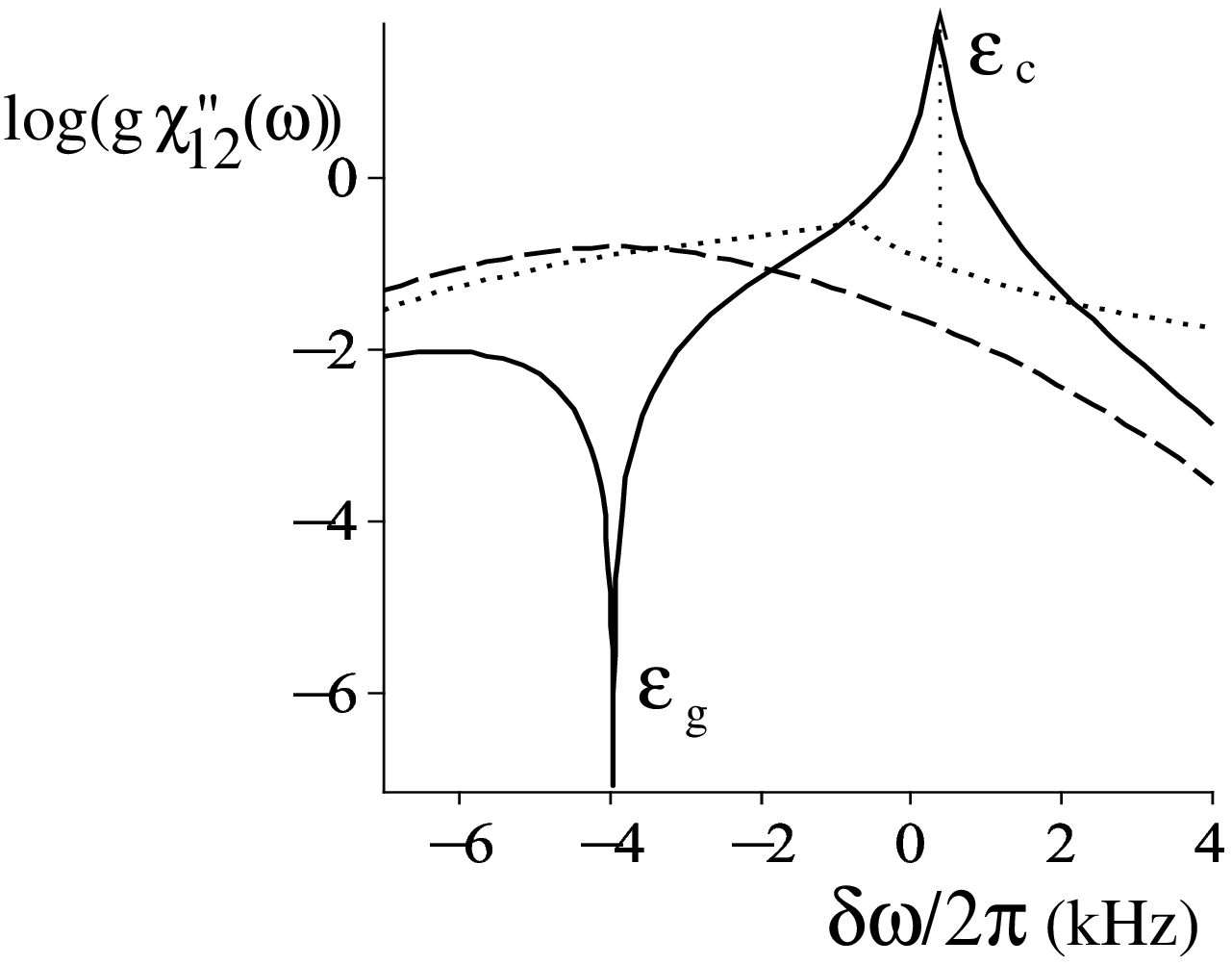}}
\caption{Idem as Fig.\ref{fig:1} but for 
$\epsilon_\vc{q}= 2\pi \times 300{\rm Hz}$. Here, the broadening of the curves is much more 
important.}
\label{fig:2}       
\end{figure}

The scattering rate is determined through the imaginary 
part of the susceptibility Eq.(\ref{chiRPA}) versus the
transition frequency $\omega$ 
and at fixed $\vc{q}$. Figs.~\ref{fig:1} and \ref{fig:2} 
show 
the corresponding  resonance around this gap in absence of screening. 
The
screening effect strongly reduces the Raman
scattering and, in particular, forbids it for 
atoms with momentum $\vc{k}$ such that $\vc{k}.\vc{q}=0$.
This case corresponds 
to $\delta \omega=\epsilon_g$ and includes 
also the condensed atoms ($\vc{k}=0$).
The graphs illustrate well the effect of the macroscopic 
wave function that deforms its shape in order to attenuate 
locally
the external potential displayed by the Raman light beams and 
to prevent incoherent scattering of the condensed atom. 
The experimental observation of this result would explain 
some of the reasons 
for which a superfluid condensate moves coherently 
without any friction with its surrounding.
Anti-screening occurs in the region close to the 
resonance frequency $\epsilon_c$ of 
the collective mode. 
At zero temperature, we recover 
$\epsilon_c=\epsilon_\vc{q}$ \cite{Fetter} while for non zero 
temperature the collective modes become damped  
for $\vc{q}\not= 0$ \cite{gap}. 

These results can be compared to the one obtained from the Bogoliubov 
non conserving approximation developed in \cite{gap} and valid only for a weakly depleted 
condensate. This approach implicitly assumes that the only elementary excitations 
are the collective ones and form a basis of quantum orthogonal states that 
describe the thermal part of the gas. Consequently, this formalism predicts 
no gap and no screening.  Instead,
the intercomponent 
susceptibility describes transitions 
involving  the two collective excitation modes of phonon 
$\epsilon_\vc{k}^B$ and of rotation in spinor space $\epsilon_\vc{k}$:
\begin{eqnarray}\label{BP}
\chi^{B}_{12}(\vc{q},\omega)=
\frac{n_\vc{0}}{\omega-\epsilon_\vc{q}+i0}+\frac{1}{V}\sum_{\pm,\vc{k}}
\frac{u^2_{\pm,\vc{k}} (n^B_{\vc{k}}+\delta_{\pm,-})}{\omega+i0 \pm
\epsilon^B_{\vc{k}}-\epsilon^{}_{\vc{k}\pm\vc{q}}}
\end{eqnarray}
where $u_{\pm,\vc{k}}=\pm
[(\epsilon_\vc{k}+gn_\vc{0})/2\epsilon^B_\vc{k}\pm 
1/2]^{1/2}$.  This function 
does not preserve the f-sum rule associated to the $SU(2)$ symmetry. In 
contrast to the GRPA, a delta peak describes a spinor rotation transition 
of the condensed fraction, and two other transitions involve the excitation transfer 
from a phonon mode into a rotation mode and the excitation creation in the two modes 
simultaneously. For small $\vc{q}$, these processes remain dispersive since 
the frequency transition depends on the momentum $\vc{k}$. As a consequence, 
the resulting spectrum shown in Figs. 5 and 6 is broader. In particular, the process 
of creation in the two modes favors transition with positive frequency. Note 
also the maximum of the curve separating the region involving a transition  
atom-atom like (high 
$\vc{k}$) and the one involving a transition phonon-atom like (low  $\vc{k}$). 

All these features established so far for the bulk case allow a clear 
comparison between the  GRPA and the Bogoliubov 
approaches. In the real case of a parabolic trap, the 
inhomogeneity induces a supplementary broadening of the spectrum 
that prevents the direct observation of the screening. This effect as well as 
the finite time resolution and the difference between the scattering lengths will 
be discussed in a subsequent work.

\section{Conclusions and perspectives}
\label{sec:3}

We have analyzed the many body properties that can be extracted 
from the Raman scattering in the framework the GRPA.
The calculated spectrum  allows to show the existence of a
second  branch of excitation but also the screening 
effect which prevents the excitation of the condensed mode alone.

The observation of  phenomena like the gap and  the 
dynamical screening could have significant
repercussions
on our microscopic understanding of a finite 
temperature Bose condensed gas 
and its superfluidity mechanism. 
On the contrary, the non-observation of these phenomena would imply
that the {\it gapless} and {\it conserving}
GRPA is not valid. In that case, a different approximation has to be developed
in order to explain what will be observed. As an alternative, 
the idea to use the  Bogoliubov approach has been also discussed. 
But unfortunately, the violation of the f-sum rule 
is a serious concern regarding this {\it non conserving} approach \cite{gap}.
All these aspects emphasize the 
importance of the experimental study of the Raman scattering at finite 
temperature.

PN thanks the referees for usefull comments  and acknowledges support 
from the Belgian FWO project G.0115.06, from the 
Junior fellowship F/05/011 of the KUL research council,  
and from the German AvH foundation.

\end{document}